\documentclass{aa}  

\usepackage{graphicx}
\usepackage{txfonts}
\newcommand\identity{1\kern-0.25em\text{l}}

\usepackage{hyperref}

\usepackage{float}

\def\Halpha{\mbox{H\hspace{0.1ex}$\alpha$}}
\def\Hbeta{\mbox{H\hspace{0.1ex}$\beta$}}

\def\Lyalpha{\mbox{Ly-\hspace{0.1ex}$\alpha$}}
\def\Lybeta{\mbox{Ly-\hspace{0.1ex}$\beta$}}
\def\Lygamma{\mbox{Ly-\hspace{0.1ex}$\gamma$}}

\def\Paalpha{\mbox{Pa-\hspace{0.1ex}$\alpha$}}

\def\CaHandK{\ion{Ca}{ii}~H\,\&\,K}

\def\CaII{\ion{Ca}{ii}}

\def\MgIb{\ion{Mg}{i}~b}

\def\MgIIHandK{\ion{Mg}{ii}~h\,\&\,k}

\def\NaId{\ion{Na}{i}~D}

\titlerunning{Markovian description of the multi-level source function}

\begin{document}

   \title{A Markovian description of the multi-level source function and its application to the Lyman series in the Sun}

   \author{K. Krikova\inst{1,2}, T. M. D. Pereira \inst{1,2}} 

   \institute{Rosseland Centre for Solar Physics, University of Oslo, PO Box 1029 Blindern, 0315 Oslo, Norway \\
   \email{kilian.krikova@astro.uio.no}
    \and
    Institute of Theoretical Astrophysics, University of Oslo, PO Box 1029, Blindern 0315, Oslo, Norway}


 
  \abstract
   {} 
   {We introduce a new method to calculate and interpret indirect transition rates populating atomic levels using Markov chain theory. Indirect transition rates are essential to evaluate interlocking in a multi-level source function, which quantifies all the processes that add and remove photons from a spectral line. A better understanding of the multi-level source function is central to interpret optically thick spectral line formation in stellar atmospheres, especially outside local thermodynamical equilibrium (LTE).}
   {We compute the level populations from a hydrogen model atom in statistical equilibrium, using the solar FALC model, a 1D static atmosphere. From the transition rates, we reconstruct the multi-level source function using our new method and compare it with existing methods to build the source function. We focus on the Lyman series lines and analyze the different contributions to the source functions and synthetic spectra.}
   {Absorbing Markov chains can represent the level-ratio solution of the statistical equilibrium equation and can therefore be used to calculate the indirect transition rates between the upper and lower levels of an atomic transition. Our description of the multi-level source function allows a more physical interpretation of its individual terms, particularly a quantitative view of interlocking. For the Lyman lines in the FALC atmosphere, we find that interlocking becomes increasingly important with order in the series, with \Lyalpha\ showing very little, but \Lybeta\ nearly 50\% and \Lygamma\ about 60\% contribution coming from interlocking. In some cases, this view seems opposed to the conventional wisdom that these lines are mostly scattering, and we discuss the reasons why.}
   {Our formalism to describe the multi-level source function is general and can provide more physical insight into the processes that set the line source function in a multi-level atom. The effects of interlocking for lines formed in the solar chromosphere can be more important than previously thought, and our method provides the basis for further exploration.}

   \keywords{ Radiative transfer -
              Line: formation --
              Methods: analytical --
              Methods: numerical --
              Sun: photosphere --
              Sun: chromosphere
               }

   \maketitle


\section{Introduction}
\label{sec:Intro}

The last decades have seen dramatic progress in numerical methods to solve the multi-level radiative transfer problem outside local thermodynamical equilibrium (LTE), so-called non-LTE methods. Starting with the pioneering work of Auer \& Mihalas \cite{Auer1969a, Auer1969b} with relaxing the assumption of LTE \citep{Holweger1967} and allowing for non-local effects in space by a two-level modeling simplification with active Lyman and Balmer continua. Followed by the successor, ``Pandora stars'' \citep{Avrett1992} allowing for non-locality in wavelength, including cross-talk between spectral lines and continua. And up to modern 1D and 3D dynamic simulations that can include the effect of non-locality in time due to the non-equilibrium ionization \citep[e.g.][]{Carlsson2002} to understand the formation of spectral lines originating from the different layers in the solar atmosphere. 

Nowadays, the ``Pandora stars'' are overtaken by modern state-of-the-art 3D radiative magneto-hydrodynamic (rMHD) simulations \citep{Gudiksen2011, Voegler2005, Rempel2017, Freytag_2013, Wray_2015, Modestov_2024, Iijima_2015} representing more realistic and detailed atmospheres. The current practice to understand spectra from the dynamic solar atmosphere is to use 3D rMHD simulations in combination with non-LTE radiative transfer codes \citep{Uitenbroek2001, Pereira2015, Carlsson1986, Leenaarts2009, Stephan_2013, Gerber_2023} to forward model spectral lines. Once the synthetic profiles are obtained it is essential to understand the basic formation mechanism to confront models with observations. To infer the formation mechanism of a spectral line, one has to investigate its source function, the ratio of the local emissivity to extinction coefficient. The source function is a key quantity in optically thick line formation that describes the weighted local addition of new photons. In modern non-LTE multi-level radiative calculations this information is encoded in the so-called ``multi-level source function'' \citep{Jefferies1968, Canfield1971, Rutten2021}.
 
The multi-level source function is obtained in terms of ratios of atomic level populations. These level ratios can be given in a general form as the ratio between direct and indirect transition rates between levels. The computation of indirect transition rates, needed for evaluating the multi-level source function, is seldom discussed in the literature. One of the most comprehensive methods for assessing indirect transition rates and probabilities was introduced by \citet{Jefferies1968}. However, the physical interpretation of indirect transition probabilities and their origin is not fully explored yet. In this paper, we present a novel approach using Markov chain theory to calculate and interpret indirect transition rates, probabilities, and the multi-level source function. A similar approach was used before by \citet{Kastner1980_a, Kastner1980_b, Kastner1982} to obtain level populations to compute intensities from optically thin lines, but not to interpret the optically thick multi-level source function. The Markov chain approach will help to interpret optically thick line formation from modern multi-level non-LTE codes and to better quantify the effects of interlocking,  which we illustrate with an application to the Lyman series in the solar atmosphere.

The outline of the paper is as follows. In Sect. \ref{sec:Methods} we introduce three different methods for calculating the multi-level source function and introduce some basic equations used in non-LTE radiative transfer. Section \ref{sec:Results} illustrates the use of the Markov chain multi-level source function description on the spectral lines \Lyalpha, \Lybeta, and \Lygamma. We present our discussion in Sect. \ref{sec:Discussion}, followed by our concluding remarks in Sect.~\ref{sec:Conclusion}.

\section{Methods}
\label{sec:Methods}

\subsection{Radiative transfer}
\label{sec:Radiative transfer}

The statistical equilibrium and radiative transfer equation are two fundamental equations in solving the non-LTE radiative problem.
The transfer equation describes how radiation is emitted, absorbed, and transported through a medium. 
In contrast, the equations of statistical equilibrium describe how atomic levels are populated under the influence of a radiation field and collisional rates. 
Solving these equations simultaneously is the main ingredient in computing synthetic spectra.

The transfer equation can be written as:
\begin{equation}\label{eq:transfer_equation}
    \frac{\textrm{d}I_\nu}{\textrm{d}s} = j_\nu - \alpha_\nu I_\nu,
\end{equation} 
with $s$ the distance measured along the beam, $j_\nu$ the emissivity, $\alpha_\nu$ the extinction coefficient, and $I_\nu$ the intensity. This equation expresses the local addition or subtraction of photons into or out of the beam. A more compact way to express the transfer equation is obtained by expressing it in terms of the optical depth $\tau_\nu$ and the source function $S_\nu \equiv j_\nu / \alpha_\nu$: 
\begin{equation}\label{eq:transfer_equation2}
    \frac{\textrm{d}I_\nu}{\textrm{d}\tau_\nu} = I_\nu - S_\nu.
\end{equation} 
The source function is a key quantity in optically thick line formation. For a spectral line, and assuming complete redistribution (CRD), it can be expressed as:
\begin{equation}\label{eq:source_function_CRD}
    S^l_{\nu_0} = \frac{2 h \nu_0^3}{c^2} \frac{1}{\frac{g_u n_l}{g_l n_u} - 1},
\end{equation} 
where $n_u$ and $n_l$ are respectively the upper and lower level populations, and $g_u$ and $g_l$ the statistical weights. The ratio between upper and lower populations is the critical quantity that varies along the atmosphere. It is typically obtained by assuming statistical equilibrium (i.e., that the populations are constant in time) and solving the system of equations:
\begin{equation}\label{eq:stat_equil}
     \sum_{j \neq i} n_j \, P_{ji} - n_i \sum_{j \neq i} P_{ij} = 0,
\end{equation}
where the first and second terms describe the rates out and into a level, respectively. $P_{ij}$ and $P_{ji}$ are the total transition rates between energy levels, which include collisional and radiative rates.

The level-ratio solution to the statistical equilibrium equations can be expressed as the ratio between direct and indirect transition rates by:
\begin{equation}
     \frac{n_l}{n_u} = \frac{P_{ul} + \sum_u}{P_{lu} + \sum_l} \label{eq:lvl_ratio_sol},
\end{equation}
where $P_{ul}$ and $P_{lu}$ are the rates for direct transitions from upper to lower and lower to upper levels, respectively. Indirect transitions are atomic transitions from the upper/lower to the lower/upper level via intermediate atomic levels and are referred to as interlocking, or multi-level detours \citep{Rutten2021}. 
These indirect transition rates are contained in the variables $\sum_u$ and $\sum_l$.
$\sum_u$ describes the indirect transition rates from the upper level to the lower level through all ``non-recurrent'' paths available in the atomic level structure.  
Sections \ref{sec:markov_msf} and \ref{sec:Lyman series} cover the meaning and calculation of all ``non-recurrent'' paths using a Markov chain.
$\sum_l$ describes the indirect rates starting from the lower level.
The indirect transition rates include information about the strength of interlocking between an atomic transition and all other transitions in an atom.

Using the level ratio solution of the statistical equilibrium equations given by Eq. (\ref{eq:lvl_ratio_sol}) we can express the line source function more intuitively.
This form is often referred to as the multi-level source function, and it includes the effects of indirect atomic transitions.
We can write a general expression for the multi-level line source function by decomposing it into three distinct contributions:
\begin{equation}
        S^l_{\nu_0} = \sigma \, \overline{J}_{\nu_0}  + \epsilon \, B_{\nu_0} (T_\mathrm{e}) + \eta \, B_{\nu_0} (T^{\star}) \label{eq:msf_sf},
\end{equation}
with
\begin{gather}
     \alpha^l_\nu = \alpha^s_\nu + \alpha^a_\nu + \alpha^d_\nu \label{eq:msf_alpha_l},  
     \\
     \sigma = \alpha^s_\nu / \alpha^l_\nu \label{eq:msf_sigma},
     \\
     \epsilon = \alpha^a_\nu / \alpha^l_\nu \label{eq:msf_epsilon},
     \\
     \eta = \alpha^d_\nu / \alpha^l_\nu \label{eq:msf_eta},
\end{gather}
where $\overline{J}_{\nu_0}$ indicates the mean radiation field, $B_{\nu_0} (T_\mathrm{e})$ a Planck function described by the electron temperature $T_e$, and $B_{\nu_0} (T^{\star})$ the interlocking source function described by characteristic temperature $T^\star$. The total line extinction coefficient $\alpha^l_\nu$ is divided into the scattering extinction $\alpha^s_\nu$, thermal extinction $\alpha^a_\nu$, and interlocking or detour extinction $\alpha^d_\nu$.
Interlocking or detour conversion refers to the same physical mechanism; throughout this paper, we use the term interlocking.

In this form of the multi-level source function, $\sigma$ describes the probability of photon extinction by scattering, $\epsilon$ the probability of photon extinction by collisional destruction, and $\eta$ the probability of photon extinction by interlocking. 

Neglecting the third term on the right side of Eq. (\ref{eq:msf_sf}) reduces the multi-level source function to the well-known form of a two-level atom. 
These two terms cover direct transitions between the upper and lower levels of a transition. 
The term $\sigma \, \overline{J}_{\nu_0}$ describes the source of photons being added to the beam from scattering.
The second term $\epsilon \, B_{\nu_0} (T_\mathrm{e})$ describes the source of photons into the beam created by collisions with electrons coming from the thermal pool.
An intuitive way to think about these two different terms is given by \cite{Rutten2021} introducing the terminology of local in space, non-local in space, and non-local in wavelength.
In this context, the thermal part describes photons created locally in space carrying information about the local plasma conditions fully described by the temperature.
The scattering part describes the effect of non-locality in space, where thermally created photons coming from different parts of the atmosphere affect the local state of the ensemble of atoms.

The third term in Eq. (\ref{eq:msf_sf}) describes the effect of a multi-level atom onto the upper and lower levels of a transition, commonly referred to as interlocking.
The interlocking term contains radiative and collisional transitions connected to the indirect transitions.
Therefore, interlocking can cover all three terminologies mentioned above namely, locality in space, non-locality in space, and non-locality in wavelength described by $\eta$ and $B_{\nu_0} (T^{\star})$.
The three different sources of extinction in Eq. (\ref{eq:msf_alpha_l}) and the interlocking source function $B_{\nu_0} (T^{\star})$ can be given as:
\begin{gather}
     \alpha^s_\nu = A_{ul} \label{eq:alpha_s},  
     \\
     \alpha^a_\nu = C_{ul} \left(1 - \mathrm{exp}\left[-\frac{h\nu_0}{k_\mathrm{B} T_\mathrm{e}}\right]\right) \label{eq:alpha_a},
     \\
     \alpha^d_\nu = \left( \sum_{u} - \frac{g_l}{g_u} \sum_{l} \right) \label{eq:alpha_d}, 
     \\
     B_\nu(T^\star) = \frac{2h\nu^3}{c^2} \left(\frac{g_u \sum_u}{g_l \sum_l} - 1 \right)^{-1}  \label{eq:eq_bstar},
\end{gather}
where $A_{ul}$, $C_{ul}$ are respectively the Einstein coefficients for spontaneous deexcitation and collisional deexcitation, and the terms $\sum_u$ and $\sum_l$ are the total transition rates from the upper/lower to the lower/upper level via all intermediate levels $t$, respectively. 

The interlocking extinction $\alpha^d_\nu$ represents the departure from a two-level atom source function. 
It illustrates that if the indirect transition rate from the upper to the lower level equals the indirect transition rate from the lower to the upper level the multi-level source function approaches the two-level source function.

%
Not many formalisms are available in the literature to evaluate the multi-level source function for model atoms with many energy levels.
The more comprehensive so far are given by \citet{Jefferies1960} and \citet{White1961}.
Here, we use a different approach to evaluate the multi-level source function, using Markov chain theory to get further insight into how levels are populated indirectly. 
In addition, we present two other formalisms in Sections \ref{sec:ETLA msf} and \ref{sec:Jefferies msf}, to compare with. 

\subsection{Equivalent two-level atom multi-level source function}
\label{sec:ETLA msf}

The most straightforward way to determine the importance of interlocking in the multi-level source function (Eq. [\ref{eq:msf_sf}]) is to make use of the level populations calculated by solving statistical equilibrium (e.g. as an output from non-LTE radiative transfer codes). 
Instead of analytically removing the level populations and expressing the level ratio solution only in terms of transition rates, one can use the statistical equilibrium equations directly.
We refer to this method as the equivalent-two-level-atom approach (ETLA) method because of its close connection to the ETLA method used in the older generations of non-LTE transfer codes, such as ``Pandora'' \citep{Avrett1992}.
The idea is to rewrite the statistical equilibrium equations in a way that direct and indirect terms of a transition are grouped:
\begin{gather}
     n_l \left( P_{lu} + a_1 \right) = n_u P_{ul} + a_2 \label{eq:etla_lower},
     \\
     n_u \left( P_{ul} + a_3 \right) = n_l P_{lu} + a_4 \label{eq:etla_upper},
     \\
     a_1 = \sum_{t \neq l,u} P_{lt} \label{eq:etla_a1},
     \\
     a_2 = \sum_{t \neq l,u} \left( n_t P_{tl} \right) \label{eq:etla_a2},
     \\
     a_3 = \sum_{t \neq l,u} P_{ut} \label{eq:etla_a3},
     \\
     a_4 = \sum_{t \neq l,u} \left( n_t P_{tu} \right) \label{eq:etla_a4},
\end{gather}
with $u$ and $l$ referring to the upper and lower levels of the transition. $P_{lu}$ and $P_{ul}$ are the direct transition rates between upper and lower levels.
Equations (\ref{eq:etla_a1})-(\ref{eq:etla_a4}) describe the indirect terms containing the radiative and collisional rates from all transient levels $t$ (excluding the upper and lower level) into and out of the upper and lower levels.
The $a_i$ terms contain the information on how strongly a transition is affected by interlocking.

Eqs. (\ref{eq:etla_lower}) and (\ref{eq:etla_upper}) can be solved in terms of level ratios where the indirect transition rates can be written as:
\begin{gather}
     \sum_u = \frac{a_2 a_3}{a_2 + a_4} \label{eq:lvl_ratio_sumu},
     \\
     \sum_l = \frac{a_1 a_4}{a_2 + a_4} \label{eq:lvl_ratio_suml},
\end{gather}
which can be used to evaluate the multi-level source function.

\subsection{Jefferies multi-level source function}
\label{sec:Jefferies msf}

Using the multi-level source function formalism by \citet{Jefferies1960, Jefferies1968} the solution of the statistical equilibrium in terms of level ratios can be given as:
\begin{equation}\label{eq:jef_level_ratio}
    \frac{n_l}{n_u} = \frac{P_{ul} + \sum_{t \neq l} P_{ut} \, q_{tl,u}}{P_{lu} + \sum_{t \neq u} P_{lt} \, q_{tu,l}},
\end{equation}
where the index $t$ refers to all transient levels between the upper and lower levels of a transition. The indirect rates $\sum_u$ and $\sum_l$ are given by:
\begin{gather}
     \sum_u = \sum_{t \neq l} P_{ut} \, q_{tl,u},
     \\
     \sum_l = \sum_{t \neq u} P_{lt} \, q_{tu,l},
\end{gather}
with the indirect transition rates not explicitly involving the level population and introducing $q_{tl,u}$ and $q_{tu,l}$, the indirect transition probabilities.
$q_{tl,u}$ describes the probability that a transition from the transient level $t$ arrives on the lower level before the upper level, and $q_{tu,l}$ in the opposite direction.
The evaluation of $q_{tl,u}$ and  $q_{tu,l}$ can be found in \cite{Jefferies1968}, and involves solving a set of linear equations, similar to the statistical equilibrium equations, to determine the probabilities. These two terms appear relatively abstract, and in the next section, we use a different approach to get insight into the indirect transition probabilities. 

\subsection{Markov chain multi-level source function}
\label{sec:markov_msf}

The motivation behind using Markov chain theory is that once we establish a connection to Jefferies's indirect transition probabilities or the analytical level ratio solution we can use all of the tools available in Markov chain theory to get insight into how levels are statistically populated given some transition rates between levels.

We start by briefly introducing the basic concepts of a Markovian process or Markov chains.
A Markovian process is an independent, finite stochastic process where the probability of going from one level to another level does not depend on the past.
By the past, we mean that the probability $p_{ij}(n)$ going from state $i$ to state $j$ after the $n$-th step does not depend on how the Markov chain reached state $i$ nor on the step count $n$.
In Markov chain theory, the terminology of states is used. Here, we will refer to states as ``levels'', as we are dealing with atomic energy levels.
Further, we indicate transition probabilities with a lowercase $p$ and transition rates with an uppercase $P$.

The statistical equilibrium equations use the transition rates $P_{ij}$ and $P_{ji}$, given per second in and out of levels $i$ and $j$.
In the Markov chain description, we transform the transition rates into transition probabilities by:
\begin{equation}\label{eq:transition_probability}
     p_{ij} = P_{ij}/ \sum_i P_{ij} = P_{ij}/ P_{i},
\end{equation}
where $P_i$ is the total rate out of level $i$.

 \begin{figure}
    \centering
    \includegraphics[width=0.5\textwidth]{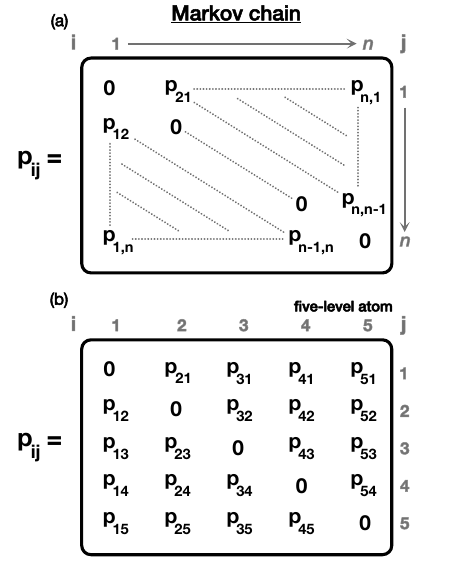}
    \caption{Markov chain transition probability matrix for a $n$-level (a) and five-level atom (b) with indicated transition probabilities $p_{ij}$.}
    \label{fig:markov_matrix}
\end{figure}
A convenient way to represent a Markov chain process is to arrange the transition probabilities between levels as a matrix. We illustrate the transition probability matrix of a $n$-level atom and five-level atom in Fig.~\ref{fig:markov_matrix}. Each column represents the transition probabilities from level $i$ into all other levels $j$.
The indices $i$ and $j$ refer to the atomic level and ``not'' the matrix rows and columns indices. If one were to exchange matrix columns, the indices would change.

To analyze a Markov chain process using a probability matrix one has to choose an initial probability vector and multiply it repeatedly with the probability matrix. The probability that the process will be in level $n$ after $k$ steps is given by
\begin{equation}\label{eq:markov_chain_process}
    \pi_k = p^2 \pi_{k-1} = p \pi_{k-2} = p^k \pi_0,
\end{equation}
where $\pi_0$ is the initial probability vector (its length depends on the model atom size), and $p^k$ contains the probabilities $p_{ij}^n$ that the Markov chain is be found in level $j$ after $n$-steps, when starting from level $i$.

 \begin{figure}
    \centering
    \includegraphics[width=0.5\textwidth]{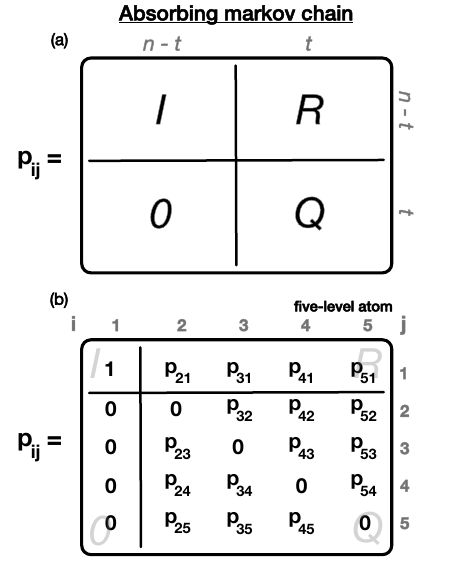}
    \caption{Absorbing Markov chain transition probability matrix for a $n$-level (a) and five-level atom (b). Panel (a) illustrates the absorbing Markov chain transition probability matrix in canonical form for $n$-level atom. The submatrices $I$ and $0$ represent the identity and zero matrix, respectively related to the absorbing level. Submatrix $R$ contains the transition probabilities from the transient level into the absorbing level. The submatrix $Q$ contains the transition probabilities between transient level. $n$ indicates the number of atomic level whereas $t$ indicates the number of transient level. Panel (b) illustrates an absorbing Markov chain transition probability matrix for a five-level atom with indicated transition probabilities $p_{ij}$.}
    \label{fig:markov_absorption_matrix}
\end{figure}
We are now interested in a special form of the probability matrix used for absorbing Markov chains.
In an absorbing Markov chain,  once the absorbing level is reached it is not possible to leave the level again.
We can transform the probability matrix shown in Fig. \ref{fig:markov_matrix}a into an absorbing Markov chain by modifying and exchanging columns.
The general idea is depicted in Fig. \ref{fig:markov_absorption_matrix}a, and involves splitting the probability matrix into four submatrices: $I$, $R$, $Q$, and a zero matrix $O$. 
$I$ represents a $(n-t) \times (n-t)$ identity matrix that contains the possible transitions after reaching the absorbing level (or levels).
The label $t$ refers to the number of transient/intermediate levels.
Figure \ref{fig:markov_absorption_matrix}b illustrates an absorbing Markov chain probability matrix for a five-level atom.
In this example, the absorbing level is the ground level, and therefore the identity matrix has only one element with $p_{11} = 1$.
The transition probabilities $p_{1j}$ are set to zero ($O$ matrix), which implies that if the absorbing Markov chain reaches the ground level, it can no longer leave it. 
$R$ concerns the transition from transient levels to absorbing levels, and has a size of $t \times (n-t)$ matrix $R$.
For the five-level atom example, the $R$ matrix is a row vector because it has only one absorbing level.
The row vector consists of all transition probabilities into the absorbing level.
The last of the four matrices is the $t \times t$ matrix $Q$. It consists of the transition probabilities between the transient levels with zeros in the diagonal.

To simulate a Markov chain process one has to calculate the powers of the transition probability matrix $p^k$ as shown in Eq. (\ref{eq:markov_chain_process}). We are interested in the case where the power $k$ is sufficiently high so that it is close enough to the limiting matrix $p^\infty_{ij}$. This limit case is illustrated in Fig. \ref{fig:markov_absorption_matrix_powers}, which can be split into four submatrices: one identity matrix, two zero matrices, and a $(I-Q)^{-1}R$ matrix that contains the probabilities starting from the transient level $n_i$ ending up in the absorbing level $n_j$. The $(I-Q)^{-1}$ submatrix contained in $p^\infty_{ij}$ is called the fundamental matrix, and is defined as:
\begin{equation}\label{eq:fundamental_matrix}
    \{N_{ij}\} = N = (I - Q)^{-1},
\end{equation}
where the $N_{ij}$ are the mean number of times a process is in the transient level $n_j$ before absorption starting from the transient level $n_i$.
Multiplying the fundamental matrix $N$ with the transition probabilities from the transient levels to the absorbing level given by $R$ gives the absorption probabilities $B_{ij}$,
\begin{equation}\label{eq:transition_prob_markov}
    \{B_{ij}\} = B = R \, N,
\end{equation}
where $B_{ij}$ are the probabilities that an absorbing chain is absorbed in level $n_j$ starting from the transient level $n_i$.
Next, we illustrate how the probabilities $B_{ij}$ look for a five-level absorbing Markov chain as shown in Fig. \ref{fig:markov_absorption_matrix}.
For a five-level atom with $n=1$ as absorbing level, the absorbing probabilities $B_{21}$ can be written after grouping terms as:
 \begin{figure}
    \centering
    \includegraphics[width=0.5\textwidth]{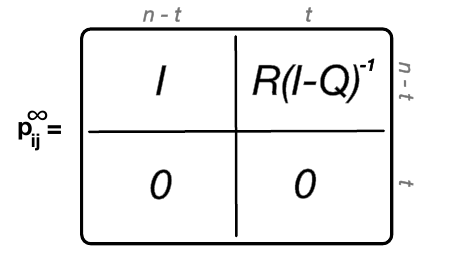}
    \caption{Limiting matrix $p^\infty_{ij}$ for a $n$-level absorbing Markov chain. The submatrices $I$ and $0$ represent the identity and zero matrix, respectively. $(I-Q)^{-1}$ represents the fundamental matrix $N$. The absorbing probabilities $B_{ij}$ are contained in the submatrix $R(I-Q)^{-1}$.}
    \label{fig:markov_absorption_matrix_powers}
\end{figure}

\begin{gather} \label{eq:absorption_probabilites_5level}
    \begin{split}
        B_{21} &= p_{21} + (p_{23} (p_{31}(1 - p_{54}p_{45}) + p_{34}p_{41} + p_{34}p_{45}p_{51} \\
         & + p_{35}p_{51} + p_{35}p_{54}p_{41}) \\
         & + p_{24} (p_{41}(1 - p_{35}p_{53}) + p_{45}p_{51} + p_{45}p_{53}p_{31} \\
         & + p_{43}p_{31} + p_{43}p_{35}p_{51}) \\
         & + p_{25} (p_{51}(1 - p_{34}p_{43}) + p_{53}p_{31} + p_{53}p_{34}p_{41} \\
         & + p_{54}p_{41} + p_{54}p_{43}p_{31}))/\det(N)\\
         & = (R_{21}N_{22} + R_{31}N_{23} + R_{41}N_{24} + R_{51}N_{25})/\det(N), \\
    \end{split}
\end{gather}
which illustrates that the total absorption probability $B_{21}$ is the sum of the direct transition probability plus the probability of all non-recurrent paths from the transient/intermediate levels $t$ to the absorbing state $k$.
%
A non-recurrent path is defined as a path of the Markov chain that does not pass through the same level twice, excluding the negative terms.
The negative terms represent closed loops that correct the different higher-order transition paths to reach the absorbing state. 
A first-order path is a multiplication of two transition probabilities, a second-order path a multiplication of three transition probabilities, and so forth.
The number of negative correction terms depends on the size of the model atom and covers all possible closed loops inside each higher-order transition path.

One can also think of $B_{21}$ as the product of the sum of the mean occupation times $N_{t1}$ and the probability of transitioning to the absorbing level $n=1$ divided by the determinant of the fundamental matrix $N$.
The term including $R_{21}N_{22}$ represents the direct transition probability between the levels $n=2$ and $n=1$.
A general description of the total absorption probability between two levels $i$ and $j$ can be written as:
\begin{equation}\label{eq:absorption_probabilites_generall}
    B_{ij} = \sum_{t \neq j} R_{tj}N_{it}/\det(N)
\end{equation}

Next, we show that the solution of the statistical equilibrium equation in terms of level ratios can be reproduced by a modified solution of the absorption Markov chain problem.
First, we need to choose an appropriate initial probability vector $\pi_0$ that will be multiplied by the limiting matrix $p^\infty_{ij}$.
The correct choice is to use the total rate out of level $P_i$ in the $\pi_0$ vector covering the $B_{ij}$ absorption probabilities.
For the case of the $B_{21}$ probability the choice of the initial probability vector will be $\pi_0 = (0, P_2, 0, 0, 0)$.
This way we transform the transition probabilities $p_{ik}$($k$ indexing overall levels except $i$) into transition rates $P_{ik}$, needed for the solution of the statistical equilibrium equation.
The last step is to divide the absorption probabilities $B_{ij}$ by the fundamental matrix entry $N_{ii}$ that leads to the cancellation of the determinant of the fundamental matrix $N$ in Eq. (\ref{eq:absorption_probabilites_generall}) and therefore to a different denominator.
Applying the steps mentioned above on the five-level case shown in Eq. (\ref{eq:absorption_probabilites_5level}) gives the total rate from level $2$ to level $1$ connected via all intermediate/transient levels $t$:
\begin{gather} \label{eq:absorption_rates_5level}
    \begin{split}
        P_{2t,1} &= P_{21} + (P_{23} (p_{31}(1 - p_{54}p_{45}) + p_{34}p_{41} + p_{34}p_{45}p_{51} \\
         & + p_{35}p_{51} + p_{35}p_{54}p_{41}) \\
         & + P_{24} (p_{41}(1 - p_{35}p_{53}) + p_{45}p_{51} + p_{45}p_{53}p_{31} \\
         & + p_{43}p_{31} + p_{43}p_{35}p_{51}) \\
         & + P_{25} (p_{51}(1 - p_{34}p_{43}) + p_{53}p_{31} + p_{53}p_{34}p_{41} \\
         & + p_{54}p_{41} + p_{54}p_{43}p_{31}))/N_{22}\\
         & = P_2(R_{21} + (R_{31}N_{23} + R_{41}N_{24} + R_{51}N_{25})/N_{22}). \\
    \end{split}
\end{gather}
%

An inconvenient aspect of the above formulation is that one has to analytically group the terms to get to the solution presented in Eq. (\ref{eq:absorption_probabilites_5level}).
However, this can be overcome by realizing that the correct grouping of terms is already contained in the limiting matrix $p^\infty_{ij}$ for a different absorbing level $k$.
Setting the absorbing level $k$ to one, as presented above, calculates all total transitions ending in the absorbing level starting from the different transient levels.
The rows of the fundamental matrix $N$ contain simultaneously the probabilities of starting at the absorbing level $k$ and ending up in each transient level by summing over the rows of the fundamental matrix.
Therefore, one can create the correct grouping of terms as in Eq. (\ref{eq:absorption_probabilites_5level}) given by:
\begin{equation}\label{eq:absorption_probabilites_generall2}
    P_{ki,j} = P_{kj} + \sum_{t \neq j, i} P_{kt} N_{tj}/N_{jj},
\end{equation}
with the first term describing the direct transition rate and the second term the indirect transition rates.
Making use of the total transition rates $P_{ki,j}$ we can express the solution of the statistical equilibrium equation in terms of level ratios,
\begin{equation}\label{eq:markov_level_ratio}
    \frac{n_l}{n_u} = \frac{P_{ut,l}}{P_{lt,u}} = \frac{P_{ul} + \sum_{t \neq u, l} P_{ut} N_{tl}/N_{ll}}{P_{lu} + \sum_{t \neq u, l} P_{lt}, N_{tu}/N_{uu}},
\end{equation}
with the labels $l$ and $u$ indicating the lower and upper level of a transition. Comparing Eqs. (\ref{eq:jef_level_ratio}) to (\ref{eq:markov_level_ratio}) shows that the Jefferies indirect transition probabilities $q_{tl,u}$ and $q_{tu,l}$ represent the terms $N_{tl}/N_{ll}$ and $N_{tu}/N_{uu}$ in the Markov chain description. These terms portray the ratios between the mean number of times the atom is in level $l$ or $u$ before being absorbed starting from the different levels.

The total indirect rates for the Markov chain description can be written as:
\begin{gather}
     \sum_u =  \sum_{t \neq u, l} P_{ut} N_{tl}/N_{ll} = \sum_{t \neq u, l} P_{ut} q_{tl,u} \label{eq:markov_sum_u},
     \\
     \sum_l = \sum_{t \neq u, l} P_{lt} N_{tu}/N_{uu} \sum_{t \neq u, l} P_{ut} q_{tu,l}  \label{eq:markov_sum_l}.
\end{gather}
We can then make use of them to determine the contribution of each intermediate level $i$ to the interlocking source function:
\begin{equation}
     B_\nu(T^\star) = \sum_i B^{\gamma_i}_\nu(T^\star) \, \gamma_i  \label{eq:interlocking_sf_split},
\end{equation}
where
\begin{gather}
     \gamma_i = \frac{g_u P_{ui} q_{il,u} -  g_l P_{li} q_{iu,l}}{g_u \sum_u - g_l \sum_l}\label{eq:interlocking_gamma},
     \\
     B^{\gamma_i}_\nu(T^\star) = \frac{2h\nu^3}{c^2} \left(\frac{g_u P_{ui} q_{il,u}}{g_l P_{li} q_{iu,l}} - 1 \right)^{-1} \label{eq:interlocking_sf_gamma}.
\end{gather}

Higher-order transition paths connected to individual intermediate levels contribute to the interlocking source function $B_{\nu_0} (T^{\star})$ by:
\begin{gather}
     B_\nu(T^\star) = \sum_i \gamma_i \, \sum_j B^{\omega_j}_\nu(T^\star) \, \omega_j \label{eq:interlocking_sf_split_loop},
     \\
     \omega_j = \frac{g_u P_{ui} q_{j}^{il,u} -  g_l P_{li} q_{j}^{iu,l}}{g_u P_{ui} q_{il,u} -  g_l P_{li} q_{iu,l}}\label{eq:interlocking_omega},
     \\
     B^{\omega_j}_\nu(T^\star) = \frac{2h\nu^3}{c^2} \left(\frac{g_u P_{ui} q_{j}^{il,u}}{g_l P_{li} q_{j}^{iu,l}} - 1 \right)^{-1} \label{eq:interlocking_sf_omega},
\end{gather}
\begin{figure}
    \centering
    \includegraphics[width=0.5\textwidth]{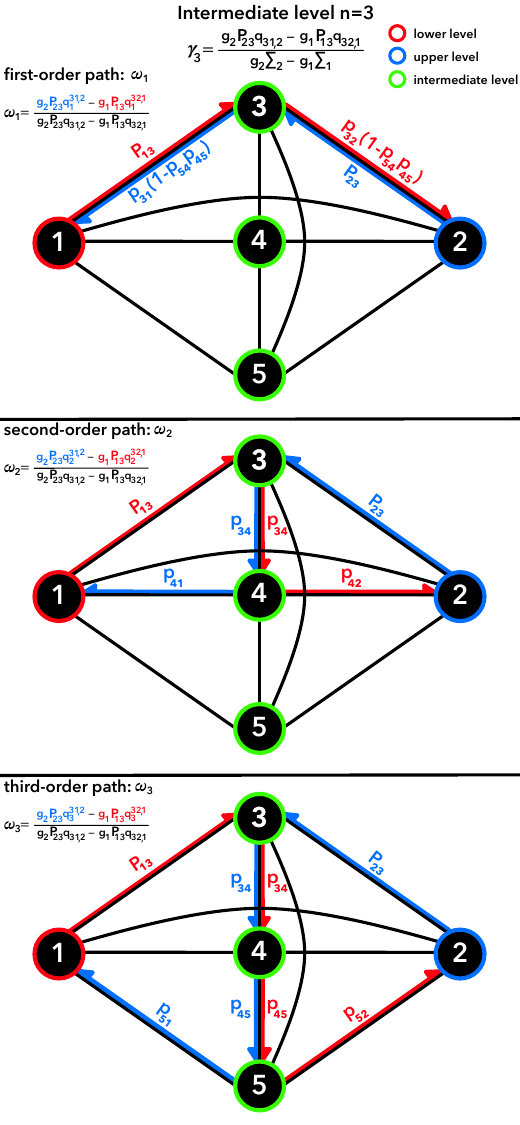}
    \caption{Possible higher-order paths through a five-level atom. Three possible paths are shown for the \Lyalpha\ transition through the intermediate level $n=3$ expressed by $\gamma_3$. The green lime color highlights the different intermediate levels. Red arrows indicate the paths from the lower level $n=1$ to the upper level $n=2$. Blue arrows indicate the paths from the upper to the lower level. \emph{Upper panel:} first-order paths indicated by $\omega_1$ with $q_{1}^{31,2}=p_{31}\left(1-p_{54}p_{45}\right)$ and $q_{1}^{32,1}=p_{32}\left(1-p_{54}p_{45}\right)$. \emph{Middle panel:} second-order paths indicated by $\omega_2$ with $q_{2}^{31,2}=p_{34}p_{41}$ and $q_{2}^{32,1}=p_{34}p_{42}$. \emph{Bottom panel:} third-order paths indicated by $\omega_3$ with $q_{3}^{31,2}=p_{34}p_{45}p_{51}$ and $q_{3}^{32,1}=p_{34}p_{45}p_{52}$.}
    \label{fig:graph}
\end{figure}
where the sum $i$ is the intermediate level and the sum $j$ is the available higher-order transition paths.

The higher-order transition paths are contained in the indirect transition probabilities $q_{iu,l}$ or $q_{il,u}$.
As an example, all possible paths for \Lyalpha\ in a five-level atom are given by Eqs. (\ref{eq:Lyalpha_q312})-(\ref{eq:Lyalpha_q512}).
The variables $q_{j}^{il,u}$ and $q_{j}^{iu,l}$ represent all possible paths connected to an intermediate level $i$ with the index $j$ summing over the order as in Eqs. (\ref{eq:Lyalpha_q312})-(\ref{eq:Lyalpha_q512}). 

We show a graphical representation of Eqs. (\ref{eq:interlocking_sf_split_loop})-(\ref{eq:interlocking_sf_omega}) in Fig. \ref{fig:graph}.
Figure \ref{fig:graph} illustrates different higher-order paths for \Lyalpha\ contained in the indirect transition probabilities $q_{31,2}$ and $q_{32,1}$ from Eq. (\ref{eq:Lyalpha_q312}) and Eq. (\ref{eq:Lyalpha_q321}).
The red paths highlight the transitions from the lower to the upper level through one or more intermediate levels. The opposite direction (upper to lower level) is shown in blue.
These represent a loop that measures the imbalance of transitions between the upper and lower levels through a given path.
Summing over all these paths available in the atomic term diagram determines the imbalance of transitions between the upper and lower levels expressed by Eq. (\ref{eq:alpha_d}).
If this imbalance is greater than the scattering extinction $\alpha_s$ or thermal extinction $\alpha_a$, interlocking effects become important and the interlocking source function dominates.

The interlocking source function $B_\nu(T^\star)$ is defined for a characteristic temperature $T^\star$, and can be split into the $B^{\gamma_i}_\nu(T^\star)$ components, which represent all paths through intermediate levels. Each $B^{\gamma_i}_\nu(T^\star)$ can be further split into $B^{\omega_j}_\nu(T^\star)$, which represent the individual paths through intermediate levels.
The weighting of the different interlocking source functions to $B_\nu(T^\star)$ is given by $\gamma_i$ and $\omega_j$.
$\gamma_i$ indicates which intermediate levels are dominant, while $\omega_j$ indicates which paths (through the dominant intermediate levels) dominate.

\subsection[Application to Lyman series]{Application to Lyman series}
\label{sec:Lyman series}

We want to demonstrate the functionality of the  Markov chain multi-level source function description introduced in Sec. \ref{sec:markov_msf} on the spectral lines \Lyalpha, \Lybeta, and \Lygamma.
With the Markov chain description of the multi-level source function, we can express the solution of the statistical equilibrium equation in terms of level ratios with Eq. (\ref{eq:markov_level_ratio}).
The level ratio for the \Lyalpha\ line can be written with algebraic indirect transition probabilities as:
\begin{gather}
      \frac{n_1}{n_2} = \frac{P_{21} + P_{23} \cdot q_{31,2} + P_{24} \cdot q_{41,2} + P_{25} \cdot q_{51,2}}{P_{12} + P_{13} \cdot q_{32,1} + P_{14} \cdot q_{42,1} + P_{15} \cdot q_{52,1}} \label{eq:Lyalpha_level_ratio}
     \\
     \begin{split}
         q_{31,2} &= (p_{31}(1 - p_{54}p_{45}) + p_{34}p_{41} + p_{34}p_{45}p_{51} + p_{35}p_{51} \\
                  &+ p_{35}p_{54}p_{41})/D \label{eq:Lyalpha_q312},
    \end{split}
     \\
     \begin{split}
         q_{41,2} &= (p_{41}(1 - p_{35}p_{53}) + p_{45}p_{51} + p_{45}p_{53}p_{31} + p_{43}p_{31} \\
                  &+ p_{43}p_{35}p_{51})/D \label{eq:Lyalpha_q412},
     \end{split}
     \\
     \begin{split}
        q_{51,2} &= (p_{51}(1 - p_{34}p_{43}) + p_{53}p_{31} + p_{53}p_{34}p_{41} + p_{54}p_{41} \\
                 &+ p_{54}p_{43}p_{31})/D \label{eq:Lyalpha_q512}
     \end{split}
    \\
     \begin{split}
         q_{32,1} &= (p_{32}(1 - p_{54}p_{45}) + p_{34}p_{42} + p_{34}p_{45}p_{52} + p_{35}p_{52} \\
                  &+ p_{35}p_{54}p_{42})/D \label{eq:Lyalpha_q321},
    \end{split}
    \\
     \begin{split}
         q_{42,1} &= (p_{42}(1 - p_{35}p_{53}) + p_{45}p_{52} + p_{45}p_{53}p_{32} + p_{43}p_{32} \\
                  &+ p_{43}p_{35}p_{52})/D \label{eq:Lyalpha_q421},
     \end{split}
    \\
     \begin{split}
        q_{52,1} &= (p_{52}(1 - p_{34}p_{43}) + p_{53}p_{32} + p_{53}p_{34}p_{42} + p_{54}p_{42} \\
                 &+ p_{54}p_{43}p_{32})/D \label{eq:Lyalpha_q521}
     \end{split}
     \\
     D = 1 - p_{34}p_{43} - p_{35}p_{53} - p_{45}p_{54} - p_{34}p_{45}p_{53} - p_{35}p_{54}p_{43} \label{eq:Lyalpha_det},
\end{gather}
for \Lybeta\ as:
\begin{gather}
      \frac{n_1}{n_3} = \frac{P_{31} + P_{32} \cdot q_{21,3} + P_{34} \cdot q_{41,3} + P_{35} \cdot q_{51,3}}{P_{13} + P_{12} \cdot q_{23,1} + P_{14} \cdot q_{43,1} + P_{15} \cdot q_{53,1}} \label{eq:Lybeta_level_ratio}
     \\
     \begin{split}
         q_{21,3} &= (p_{21}(1 - p_{54}p_{45}) + p_{24}p_{41} + p_{24}p_{45}p_{51} + p_{25}p_{51} \\
                  &+ p_{25}p_{54}p_{41})/D \label{eq:Lybeta_q213},
    \end{split}
     \\
     \begin{split}
         q_{41,3} &= (p_{41}(1 - p_{25}p_{52}) + p_{42}p_{21} + p_{42}p_{25}p_{51} + p_{45}p_{51} \\
                  &+ p_{45}p_{52}p_{21})/D \label{eq:Lybeta_q413},
     \end{split}
     \\
     \begin{split}
        q_{51,3} &= (p_{51}(1 - p_{24}p_{42}) + p_{52}p_{21} + p_{52}p_{24}p_{41} + p_{54}p_{41} \\
                 &+ p_{54}p_{42}p_{21})/D \label{eq:Lybeta_q513}
     \end{split}
    \\
     \begin{split}
         q_{23,1} &= (p_{23}(1 - p_{54}p_{45}) + p_{24}p_{43} + p_{24}p_{45}p_{53} + p_{25}p_{53} \\
                  &+ p_{25}p_{54}p_{43})/D \label{eq:Lybeta_q231},
    \end{split}
    \\
     \begin{split}
         q_{43,1} &= (p_{43}(1 - p_{25}p_{52}) + p_{42}p_{23} + p_{42}p_{25}p_{53} + p_{45}p_{53} \\
                  &+ p_{45}p_{52}p_{23})/D \label{eq:Lybeta_q431},
     \end{split}
    \\
     \begin{split}
        q_{53,1} &= (p_{53}(1 - p_{24}p_{42}) + p_{52}p_{23} + p_{52}p_{24}p_{43} + p_{54}p_{43} \\
                 &+ p_{54}p_{42}p_{23})/D \label{eq:Lybeta_q531}
     \end{split}
     \\
     D = 1 - p_{24}p_{42} - p_{25}p_{25} - p_{45}p_{54} - p_{24}p_{45}p_{52} - p_{25}p_{54}p_{42} \label{eq:Lybeta_det},
\end{gather}
and for \Lygamma\ as:
\begin{gather}
      \frac{n_1}{n_4} = \frac{P_{41} + P_{42} \cdot q_{21,4} + P_{43} \cdot q_{31,4} + P_{45} \cdot q_{51,4}}{P_{14} + P_{12} \cdot q_{24,1} + P_{13} \cdot q_{34,1} + P_{15} \cdot q_{54,1}} \label{eq:Lycont_level_ratio}
     \\
     \begin{split}
         q_{21,4} &= (p_{21}(1 - p_{35}p_{53}) + p_{23}p_{31} + p_{23}p_{35}p_{51} + p_{25}p_{51} \\
                  &+ p_{25}p_{53}p_{31})/D \label{eq:Lycont_q215},
    \end{split}
     \\
     \begin{split}
         q_{31,4} &= (p_{31}(1 - p_{25}p_{52}) + p_{32}p_{21} + p_{32}p_{25}p_{51} + p_{35}p_{51} \\
                  &+ p_{35}p_{52}p_{21})/D \label{eq:Lycont_q315},
     \end{split}
     \\
     \begin{split}
        q_{51,4} &= (p_{51}(1 - p_{23}p_{32}) + p_{52}p_{21} + p_{52}p_{23}p_{31} + p_{53}p_{31} \\
                 &+ p_{53}p_{32}p_{21})/D \label{eq:Lycont_q513}
     \end{split}
    \\
     \begin{split}
         q_{24,1} &= (p_{24}(1 - p_{35}p_{53}) + p_{23}p_{34} + p_{23}p_{35}p_{54} + p_{25}p_{54} \\
                  &+ p_{25}p_{53}p_{34})/D \label{eq:Lycont_q231},
    \end{split}
    \\
     \begin{split}
         q_{34,1} &= (p_{34}(1 - p_{25}p_{52}) + p_{32}p_{24} + p_{32}p_{25}p_{54} + p_{35}p_{54} \\
                  &+ p_{35}p_{52}p_{24})/D \label{eq:Lycont_q431},
     \end{split}
    \\
     \begin{split}
        q_{54,1} &= (p_{54}(1 - p_{23}p_{32}) + p_{52}p_{24} + p_{52}p_{23}p_{34} + p_{53}p_{34} \\
                 &+ p_{53}p_{32}p_{24})/D \label{eq:Lycont_q531}
     \end{split}
     \\
     D = 1 - p_{23}p_{32} - p_{25}p_{52} - p_{35}p_{53} - p_{23}p_{35}p_{52} - p_{25}p_{53}p_{32} \label{eq:Lycont_det}.
\end{gather}

The above level ratios for \Lyalpha, \Lybeta, and \Lygamma\ are then used to determine the dominant intermediate level in the interlocking source function given by Eq. (\ref{eq:interlocking_sf_split}).
Once the dominant intermediate level is known one can determine the dominant loop given by the algebraic expressions for the indirect transition probabilities $q_{il,u}$ and $q_{iu,l}$.
The variables $D$ represent all possible closed loops, meaning loops not connected to the upper and lower levels of the atomic transition.
With this formalism, we clearly identify which indirect transition(s) create most of the line photons observed in the spectral line of interest.

\subsection{Synthetic spectra}
\label{sec:Synthetic spectra}

We applied our method to spectral lines synthesized from the FALC solar model \citep{Fontenla1993}, a commonly used 1D plane-parallel and static model of the quiet Sun.

\begin{figure*}
    \centering
    \includegraphics[width=1\textwidth]{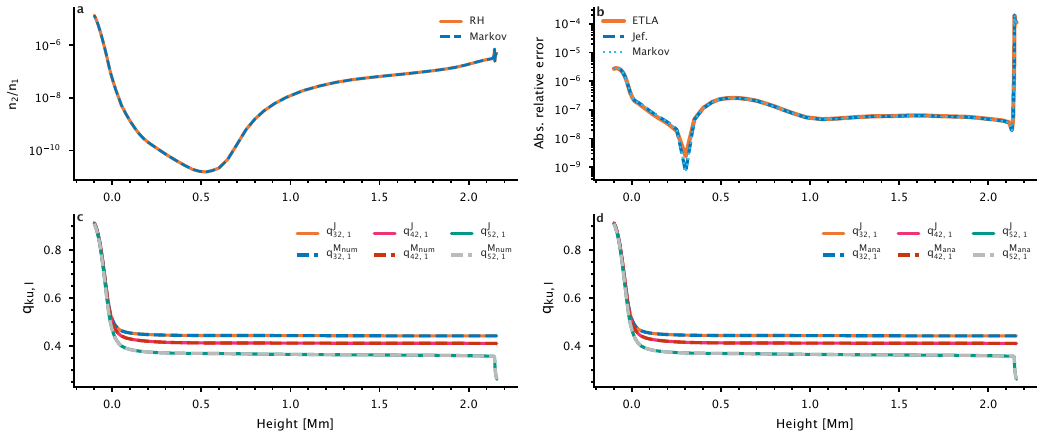}
    \caption{Accuracy of the Markov chain level ratio solution compared to the ETLA and Jefferies method. Panel (a) compares the RH  $n_2/n_1$ level ratio solution to the Markov chain solution. Panel (b) compares the absolute relative error of the ETLA, Jefferies, and Markov method against the RH solution. Panel(c) shows the indirect transition probabilities numerically calculated with the Jefferies ($q^\mathrm{J}_{i2,1}$) and Markov method ($q^{\mathrm{M}_{\mathrm{num}}}_{i2,1}$) for the $n_2/n_1$ level ratio. Panel (d) shows the analytical indirect transition probabilities from the Markov method ($q^{\mathrm{M}_{\mathrm{ana}}}_{i2,1}$) against Jefferies ($q^\mathrm{J}_{i2,1}$).}
    \label{fig:verification}
\end{figure*}

We synthesized the hydrogen Lyman series lines using the RH 1.5D code \citep{Pereira2015}. 
RH 1.5D is based on the RH code \citep{Uitenbroek2001} and solves the multilevel non-LTE radiative transfer problem with overlapping active bound-bound and bound-free transitions in 1D geometry based on the method developed in a series of papers by \citet{Rybicki1991, Rybicki1992, Rybicki1994}.

To synthesize the Lyman series lines we used a five-level hydrogen model atom (including the continuum). 
All hydrogen transitions were treated with the assumption of complete frequency redistribution (CRD) over the line profiles. 
Further, we included line blends in the Balmer continuum radiation fields in RH 1.5D in the form of a Kurucz line list \footnote{Details can be found at:  \url{http://kurucz.harvard.edu/linelists.html}}. 
We only included lines in the Kurucz line list that significantly alter the Balmer continuum radiation field \citep[line list taken from][]{Krikova2023} affecting the hydrogen ionization in the atmosphere \citep[see e.g.][]{Carlsson2002}.

From the converged solution of RH 1.5D, we extracted the transition rates for all relevant transitions, and used them to build the population ratios and the multi-level source functions using different methods.

\section{Results}
\label{sec:Results}

First, we quantified how accurate the Markov chain level ratio solution is compared to the ETLA and Jefferies method. 
In Fig. \ref{fig:verification} we compare the three different methods for the ratio $n_2/n_1$.
From Fig. \ref{fig:verification}a it is clear that the Markov chain solution for the level ratio agrees very well with the solution from RH.
The absolute relative error (Fig. \ref{fig:verification}b) between the three methods and the RH solution stays below $0.1\%$ throughout the atmosphere. The largest errors occur in the transition region.
The absolute errors from the Markov chain and Jefferies methods are similar throughout the atmosphere, but the ETLA method shows a deviation close to the temperature minimum.
We find a similar behavior for other level population ratios except that the ETLA method shows a lower error in the lower part of the atmosphere (below $0.5$~Mm).
The accuracy of the indirect transition probabilities are shown in Fig. \ref{fig:verification}c and Fig. \ref{fig:verification}d.
The Markov chain description's analytical and numerical indirect transition probabilities match those calculated with the Jefferies method.

\begin{figure*}
    \centering
    \includegraphics[width=1\textwidth]{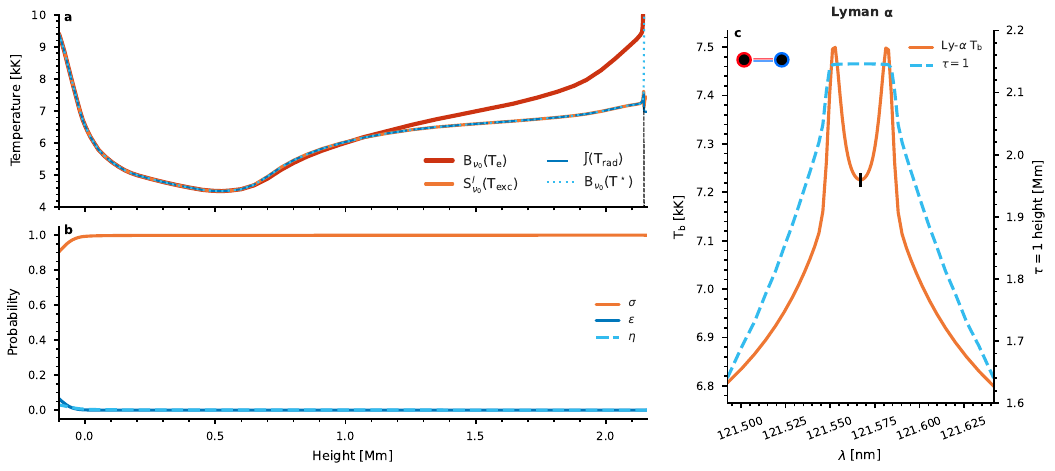}
    \caption{Multi-level source function description of the \Lyalpha\ spectral line synthesized from the FALC atmosphere. Panel (a) shows the computed \Lyalpha\ source function from RH ($S_{\nu_0}^l(T_\mathrm{exec})$) as well as the three components describing the multi-level source function, the Planck function $B_{\nu_0}(T_\mathrm{e})$, the mean radiation field $\overline{J}(T_\mathrm{rad})$, and the interlocking source function $B_{\nu_0}(T^{\star})$ in units of brightness temperature. The black dashed line (top of the atmosphere) shows the normalized contribution function to intensity for the wavelength position indicated with a black vertical in Panel (c). Panel (b) illustrates the scattering probability $\sigma$, destruction probability $\epsilon$, and interlocking probability $\eta$. Panel (c) shows the emergent line profile in units of brightness temperature as well as the optical depth unity height with the axis to the right. The icon in the top left corner indicates that two-level processes dominate the \Lyalpha\ source function.}
    \label{fig:lyalpha_mls}
\end{figure*}

\paragraph{\Lyalpha\ multi-level source function.}
%
In Fig.~\ref{fig:lyalpha_mls} we quantified interlocking on the $\Lyalpha$\ line, computed from the FALC model.
From Figure \ref{fig:lyalpha_mls}b one sees that the \Lyalpha\ source function is dominated by scattering from the photosphere to the transition region.
This shows that the $\Lyalpha$ source function is well approximated by a two-level source function. This is illustrated by a diagram in the top left corner of Fig. \ref{fig:lyalpha_mls}c.
The $\Lyalpha$ source function follows the mean radiation field, which in this case is also equal to the interlocking source function (except in the transition region).
The interlocking source function shows a strong rise in the transition region due to the sharp temperature increase.
The variation of the mean radiation field $\overline{J}_{\nu_0}$ in the atmosphere is represented in the \Lyalpha\ emergent intensity, shown in Fig. \ref{fig:lyalpha_mls}c with a central reversal.
The emergent intensity of the line profile is coming from narrow heights in the FALC atmosphere, mapping $\overline{J}_{\nu_0}$ into intensity.
The central reversal (wavelength position indicated with a black vertical line) is formed in the topmost part of the FALC atmosphere with an outward decreasing source function after an initial peak in the transition region.
This behavior of the source function gives rise to the central depression; the two peaks mark the peak of the source function in the transition region.
In summary, the \Lyalpha\ source function is non-local in space and is affected by different parts of the atmosphere above the temperature minimum. Two-level scattering is the dominant term in the multi-level source function.

\begin{figure*}
    \centering
    \includegraphics[width=1\textwidth]{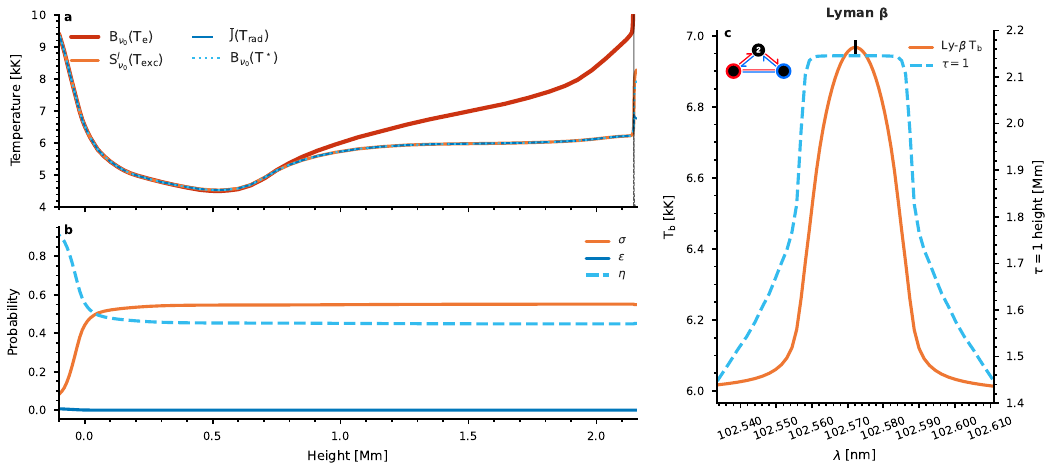}
    \caption{Multi-level source function description of the \Lybeta\ spectral line synthesized from the FALC atmosphere. Figure caption is the same as for Fig. \ref{fig:lyalpha_mls} except for the icon. The icon in the top left corner indicates that two-level processes and a first-order interlocking process through the level $n=2$ dominates the \Lybeta\ source function.}
    \label{fig:lybeta_mls}
\end{figure*}

\paragraph{\Lybeta\ multi-level source function.} In Fig. \ref{fig:lybeta_mls} we show the results for \Lybeta.
For this line, interlocking dominates below the photosphere, which thermalizes the \Lybeta\ source function to the Planck function.
Upwards, from the photosphere through the chromosphere up to the transition region, the \Lybeta\ source function becomes a combination of interlocking and scattering.
Interlocking contributes $\approx 45\%$ to the \Lybeta\ source function with the rest coming from the mean radiation field due to scattering.
The mean radiation field matches the interlocking source function throughout the FALC atmosphere except in the transition region, where the interlocking source function shows a stronger sensitivity to the temperature rise, such as for \Lyalpha.
This sensitivity of the source function to temperature is reflected in the \Lybeta\ emission profile shown in Fig. \ref{fig:lybeta_mls}c; the line core is formed in the transition region.
A significant amount of the line core photons of \Lybeta\ are created by interlocking processes, giving rise to the strong \Lybeta\ emission profile.
The dominating interlocking process is shown by the diagram in the top left corner of Fig. \ref{fig:lybeta_mls}c, indicating a first-order path through the intermediate level $n=2$.

\begin{figure}
    \centering
    \includegraphics[width=0.5\textwidth]{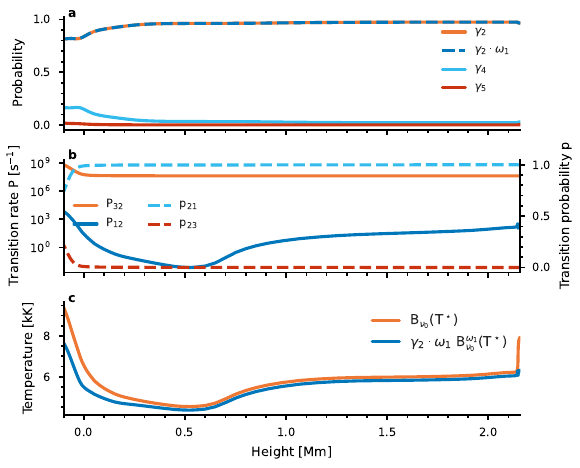}
    \caption{Interlocking source function description of the \Lybeta\ spectral line synthesized from the FALC atmosphere. Panel (a) shows the contribution to interlocking by the different intermediate states, expressed by \ref{eq:interlocking_gamma} and paths, expressed by \ref{eq:interlocking_omega}. $\gamma_2$, $\gamma_4$, and $\gamma_5$ refers to the intermediate states $n=2$, $n=4$, and continuum, respectively. $\omega_1$ refers to the first-order path connected to $p_{21}(1 - p_{54}p_{45})$ and $p_{23}(1 - p_{54}p_{45})$. Panel (b) explains the dominance of the first-order path $\omega_1$, with $P_{ij}$ the transition rates and $p_{ij}$ transition probabilities between states. Panel (c) shows the contribution of the source function connected to $\omega_1$ (blue) to the ``total'' interlocking source function.}
    \label{fig:lybeta_interlocking}
\end{figure}

Figure \ref{fig:lybeta_interlocking} displays in detail how strongly each intermediate level and path dominates the interlocking source function for \Lybeta.
From below the photosphere up to the transition region the dominant intermediate level is $n=2$, indicated with $\gamma_2$ in Fig. \ref{fig:lybeta_interlocking}a.
Indirect transitions through the levels $n=4$ ($\gamma_4$) and $n=5$ ($\gamma_5$) have a negligible contribution to the interlocking source function. The intermediate level $n=4$ has a little contribution only below the surface.

After determining the dominant intermediate level we evaluated which higher-order transition path dominates the indirect transition probabilities $q_{21,3}$ and $q_{23,1}$ highlighted in Eqs. (\ref{eq:Lybeta_q213}) and (\ref{eq:Lybeta_q231}).
Fig. \ref{fig:lybeta_interlocking}a highlights that $\gamma_2$ is dominated by a first-order interlocking process, $\omega_1$, connected to \Lyalpha\ and \Halpha.
This first-order interlocking process is described by a path connecting the upper and lower level of \Lybeta\ with the transition probabilities $p_{21}(1 - p_{54}p_{45})$ and $p_{23}(1 - p_{54}p_{45})$.
Fig. \ref{fig:lybeta_interlocking}b illustrates the cause of the strong interlocking of \Lybeta\ with \Lyalpha\ and \Halpha\ by displaying the important terms occurring in $\omega_1$.
The transition rate $P_{32}$ is dominated by the radiative rates given by the spontaneous deexcitation of \Halpha, and is therefore nearly constant through the atmosphere.
The transition rate $P_{12}$ is dominated by radiative excitation from the ground state by the mean radiation field of \Lyalpha.
Most hydrogen atoms are in the ground state. 
Therefore the radiative rates $R_{12}$ are small compared to the spontaneous deexcitation rate of \Halpha, which leads to orders of magnitude differences between $P_{32}$ and $P_{12}$.

The large difference between the transition probabilities $p_{21}$ and $p_{23}$ results from the following.
\Lyalpha\ has the highest spontaneous deexcitation rate of all hydrogen lines in the solar spectrum and is dominating the transitions out of the level $n=2$.
This is illustrated by $p_{21}$ being close to one through the FALC atmosphere whereas the probability $p_{23}$ is relatively small.
This specific combination of transition rates and probabilities results in a large imbalance in Eq. (\ref{eq:alpha_d}). This is comparable in size to the scattering extinction (Eq. [\ref{eq:alpha_s}]), making interlocking an important process for \Lybeta.
Figure \ref{fig:lybeta_interlocking}c shows the contribution of the first-order interlocking source function $B^{\omega_1}_\nu(T^\star)$ to the total interlocking source function $B_\nu(T^\star)$.

To summarize, the \Lybeta\ source function is non-local in space and wavelength.
The non-locality in wavelength comes from interlocking with \Lyalpha\ and \Halpha.
The non-locality in space above the temperature minimum stems from the fact that \Lyalpha\ and \Halpha\ are strongly scattering.

\begin{figure*}
    \centering
    \includegraphics[width=1\textwidth]{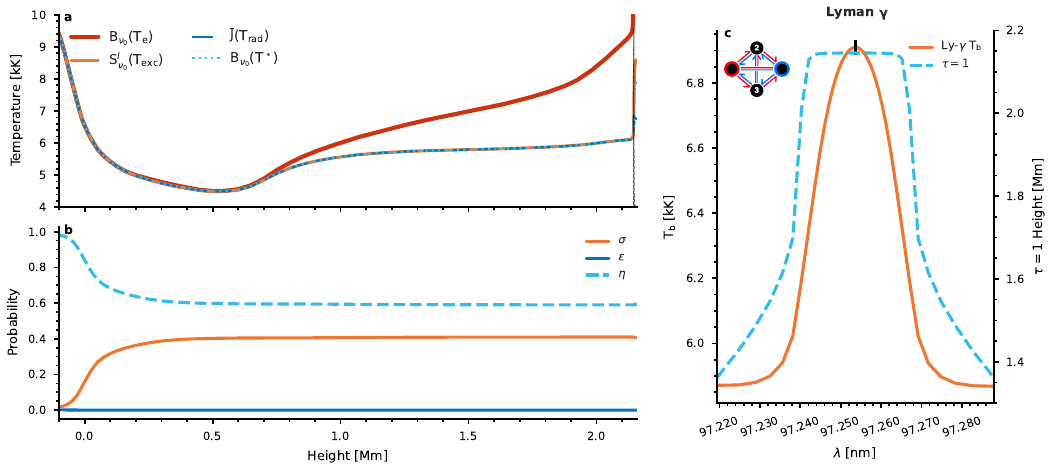}
    \caption{Multi-level source function description of the \Lygamma\ spectral line synthesized from the FALC atmosphere. Figure caption is the same as for Fig. \ref{fig:lyalpha_mls} except for the icon. The icon in the top left corner indicates that two-level processes as well as a first-order interlocking process through the level $n=2$ and a first-order and second-order interlocking process through the level $n=3$ dominate the \Lygamma\ source function.}
    \label{fig:lygamma_mls}
\end{figure*}

\paragraph{\Lygamma\ multi-level source function.} Lastly, we turn our attention to \Lygamma, which we show in Fig. \ref{fig:lygamma_mls}.
Below the photosphere, the source function is dominated by interlocking with the interlocking source function thermalized to the Planck function, similar to \Lybeta.
From the photosphere up to the transition region interlocking plays a significant role in setting the \Lygamma\ source function coupled to the temperature, until the temperature minimum.
The mean radiation field $\overline{J}(T_\mathrm{rad})$ and interlocking source function $B_{\nu_0}(T^{\star})$  both approximate well the \Lygamma\ line source function (except in the transition region).
In the transition region, the interlocking source function responds strongly to the temperature rise similar to \Lyalpha\ and \Lybeta\, where the \Lygamma\ line core is formed.
\Lygamma\ is formed from the chromosphere ($\approx 1.35$~Mm) up to the transition region ($\approx 2.15$~Mm). Its emergent intensity reflects the variation of the interlocking source function between these heights.
Next, we want to address which interlocking process sets the behavior of the \Lygamma\ source function through the FALC atmosphere.

\begin{figure}
    \centering
    \includegraphics[width=0.5\textwidth]{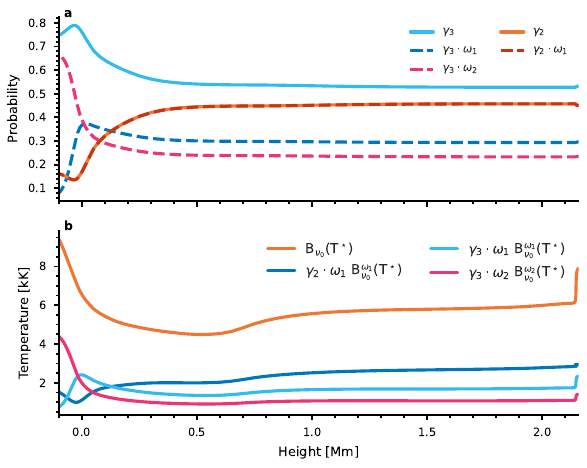}
    \caption{Interlocking source function description of the \Lygamma\ spectral line synthesized from the FALC atmosphere. Panel (a) shows the contribution to interlocking by the different intermediate states, expressed by \ref{eq:interlocking_gamma} and higher-order paths, expressed by \ref{eq:interlocking_omega}. $\gamma_2$ and $\gamma_3$ refers to the intermediate states $n=2$ and $n=3$. $\omega_1$ and $\omega_2$ connected to $\gamma_3$ refer to the first-order path $p_{31}(1 - p_{25}p_{52})$ and $p_{34}(1 - p_{25}p_{52})$ and second-order path $p_{32}p_{21}$ and $p_{32}p_{24}$, respectively. $\omega_1$ connected to $\gamma_2$ refers to the first-order path connected to $p_{21}(1 - p_{35}p_{53})$ and $p_{24}(1 - p_{35}p_{53})$. Panel (b) shows the contributions to the ``total'' interlocking source function by $\omega_1$ through the intermediate state $n=2$ as well as $\omega_1$ and $\omega_2$ through the intermediate state $n=3$.}
    \label{fig:lygamma_interlocking}
\end{figure}

In Fig. \ref{fig:lygamma_interlocking} we present the contribution to the interlocking source function from the different intermediate levels.
Fig. \ref{fig:lygamma_interlocking}a highlights that transitions through two intermediate levels control the interlocking source function.
The first intermediate level is $n=2$ ($\gamma_2$) dominated by a first-order path, while the second intermediate level $n=3$ ($\gamma_3$) is dominated by a first-order and a second-order path.
Below the photosphere, the intermediate level $n=3$ dominates the interlocking process through a second-order path described by $p_{32}p_{21}$ and $p_{32}p_{24}$ connected to the \Halpha, \Hbeta, and \Lyalpha\ transitions.
Starting from the photosphere, the first-order path through the intermediate level $n=2$ becomes increasingly important until approximately the temperature minimum.
From that point on, the contribution from intermediate levels $n=2$ and $n=3$ stay roughly the same level with the intermediate level $n=3$ slightly dominating up to the transition region.
The contribution to the total interlocking source $B_\nu(T^\star)$ function from the higher-order interlocking source functions $B^{\omega_j}_\nu(T^\star)$ from the different intermediate levels can be seen in Fig. \ref{fig:lygamma_interlocking}b.

The \Lygamma\ source function is non-local in space and wavelength, similar to \Lybeta.
The non-locality in wavelength comes from interlocking with \Lyalpha, \Halpha, \Hbeta,  \Paalpha\ affecting the \Lygamma\ source function.
The source function is local in space until the temperature minimum. Further up, the source function becomes non-local in space.
The non-locality in space comes from the scattering nature of \Lyalpha\ and the other spectral lines affecting \Lygamma.

\section{Discussion}
\label{sec:Discussion}

The Markov chain description we present in Sect. \ref{sec:markov_msf} is a different approach to interpreting a multi-level source function with level ratios, which we built by using transition rates from a converged non-LTE solution.
It can be used to calculate the indirect transition rates essential to evaluate the multi-level source function and to determine which interlocking processes are important, and where.
The indirect transition rates are expressed in terms of indirect transition probabilities as introduced by \cite{Jefferies1960}. The absorbing Markov chain approach is a different method of computing the indirect transition probabilities. It can be used numerically and analytically.
Our Markov chain approach can be used to get a deeper understanding of the indirect transition probabilities and how the terms build-up for larger model atoms.
One realizes that the indirect transition probabilities contain all paths via the intermediate level to the upper/lower without going through the same intermediate level twice (correcting for closed loops).
The Markovian description of the multi-level source function can be used to determine which physical processes dominate the source function of a spectral line: scattering, thermal, or interlocking.

We applied the Markovian description of the multi-level source function to the \Lyalpha, \Lybeta, and \Lygamma\ lines from the FALC model. 
We find that the \Lyalpha\ source function is dominated by two-level scattering due to the large spontaneous photo-deexcitation rate, orders of magnitude larger than the interlocking extinction $\alpha^d_\nu$ (Eq. [\ref{eq:alpha_d}]), which results in a low photon destruction probability for \Lyalpha\ photons \citep{Rutten2021, Rutten2017, Rutten2017_b}.
Surprisingly, the Lyman series line source functions above \Lyalpha\ start to show a significant contribution from interlocking processes.
\Lybeta\ shows a strong coupling with the \Lyalpha\ and \Halpha\ lines, seemingly in contradiction with the suggestion by \citet{Rutten2017_b} that \Lybeta\ is well approximated by two-level atom scattering.
\cite{Skumanich1986} already suggested that \Lybeta\ is strongly coupled with \Lyalpha\ and \Halpha\, and used a different approach to study the multi-level source function by applying a sensitivity analysis of the statistical equilibrium equations by perturbing the atomic transition rates.
Their study also suggested that the \Lyalpha\ source function is strongly influenced by the \Lybeta\ and \Halpha\ transitions, reflected by the variations of the \Lyalpha\ mean radiation field given by the relation
\begin{equation}\label{eq:Lyalpha_mean_rad_sku}
    \overline{J}_{\mathrm{Ly}-\alpha} \approx \frac{2h\nu_0}{c^2} \frac{\left( B \overline{J} \right)_{13}}{\left( B \overline{J} \right)_{23}} \frac{A_{32}}{A_{31}},
\end{equation}
where $A$, $B$, and $\overline{J}$ are the Einstein coefficients for spontaneous deexcitation, radiative excitation, and mean radiation fields, respectively.
%
%
%
We find that the direct radiative rates dominate the \Lyalpha\ source function following the mean radiation field.
However, looking into the \Lyalpha\ indirect transition rates we find that a first-order path through the intermediate level $n=3$ dominates the indirect transition rates connected to the atomic transitions \Halpha\ and \Lybeta.
Therefore, we find a similar relation for the \Lyalpha\ mean radiation field given by:
\begin{equation}\label{eq:Lyalpha_mean_rad_markov}
    \overline{J}_{\mathrm{Ly}-\alpha} \approx \frac{2h\nu_0}{c^2} \frac{\left( B \overline{J} \right)_{13}}{\left( B \overline{J} \right)_{23}} \frac{p_{32}}{p_{31}},
\end{equation}
which gives the same result as \citet{Skumanich1986} and is valid up to the transition region of the FALC atmosphere.
Our multi-level source function description indicates that the \Lyalpha\ radiation field is important for the formation of some hydrogen lines, as suggested by \citet{Skumanich1986}.
This is particularly true for hydrogen lines that share an upper or lower level with the \Lyalpha\ transition, such as \Lyalpha, \Lybeta, and \Halpha.

An unexpected coincidence revealed by our approach is that the mean radiation field $\overline{J}(T_\mathrm{rad})$ and interlocking source function $B_{\nu_0}(T^{\star})$ are equal for \Lyalpha, \Lybeta, and \Lygamma\ up to the transition region in the FALC atmosphere.
This needs a more detailed explanation.

We base our explanation on Eq. (\ref{eq:lvl_ratio_sol}), which expresses the level ratio solution regarding direct and indirect transition rates between levels.
In the multi-level source function description, the ratio of the direct terms ${n_l}/{n_u} = {P_{ul}}/{P_{lu}}$ represents the two-level term $ \sigma \, \overline{J}_{\nu_0}  + \epsilon \, B_{\nu_0} (T_\mathrm{e})$, whereas the ratio of the indirect terms ${n_l}/{n_u} = {\sum_u}/{\sum_l}$ represents the interlocking term $\eta \, B_{\nu_0} (T^{\star})$.
If the direct transition rates dominate, $P_{ul} \gg \sum_u$ and $P_{lu} \gg \sum_u$; the multi-level source function is reduced to the two-level approximation (e.g. as with \Lyalpha).
If the indirect transition rates $\sum_u$ and $\sum_l$ become comparable in size to the direct transition rates, interlocking processes become important (as illustrated by \Lybeta\ and \Lygamma).
For the mean radiation field $\overline{J}_{\nu_0}$ to be equal the interlocking source function $B_{\nu_0}(T^{\star})$  we need
\begin{equation}\label{eq:two_lvl_detailed_balance}
   \frac{n_l}{n_u} \approx \frac{P_{ul}}{P_{lu}} \approx \frac{\sum_u}{\sum_l}.
\end{equation}
This implies that the spectral line of interest has to be formed under a ``two-level'' or ``interlocking'' detailed balance.
There are as many upward as downward transitions in direct and indirect transitions between the upper and lower levels of an atomic transition.

To explain the implications of Eq. (\ref{eq:two_lvl_detailed_balance}) we draw an analogy to coherent and non-coherent scattering in spectral lines.
Coherent scattering assumes no redistribution in the frequency of a line photon after each scattering event; therefore, each frequency position in the spectral line is independent of all other frequency positions in the spectral profile.
No cross-talk between photons at different frequencies is allowed.
In the case of non-coherent scattering, photons redistribute over the entire line profile \citep{Jefferies1968}.
An analogy can be drawn to the effect of interlocking.
If a multi-level atom can be approximated by a two-level atom, the line photons arise only from processes related to the upper and lower level of a transition.
However, if interlocking becomes important line photons can arise from any transition in the atom, thereby making all transitions a potential photon source for the given spectral line.
The mean radiation field of a particular transition becomes a function of the mean radiation field of one or more transitions of an atom, as indicated by Eq. (\ref{eq:Lyalpha_mean_rad_markov}).
Therefore, if the indirect transition rates become comparable to the direct transition rates, interlocking can have a strong influence on the mean radiation field of a line, leading to a two-level detailed balance.

The effect of a two-level detailed balance can lead to a misinterpretation of the physical mechanism dominating the line source function.
One might conclude that the line source function is represented by the two-level approximation very accurately, when in fact the indirect transition rates dominate over the direct transition rates.
Therefore, one should always compute the indirect transition rates $\sum_u$ and $\sum_l$ and compare them to the direct transition rates.
\Lybeta\ and \Lygamma\ are great examples where a two-level source function approximation would lead to an incorrect classification of the physical mechanism dominating the line source function.

The multi-level source function we introduce in Sect. \ref{sec:markov_msf} is valid only under CRD.
The more general assumption of PRD over the line profile would result in a multi-level source function that includes the different wavelength-dependent line profiles for emission, absorption, and stimulated emission.
The PRD line source function would become wavelength-dependent and more applicable for resonance lines like \Lyalpha.
However, the dominant physical mechanism dominating the line source should not change because the level ratios are wavelength-independent.
It would mainly depend on how strongly the assumption of PRD changes the mean radiation fields connected to the different spectral lines influencing the transition probabilities used for the absorption Markov chain.
One could include some effects of PRD by assuming that the stimulated emission and absorption profiles are equal $\phi_{lu} = \chi_{ul}$. 
This makes the source function wavelength dependent and one can write the lines source function given in Eq. (\ref{eq:msf_sf}) as:
\begin{equation}
   S_\nu^{\mathrm{PRD}} = S^\mathrm{l}_{\nu_0} \frac{\psi_{ul}}{\phi_{lu}},
\end{equation}
partly including PRD effects. 

Many spectral lines formed in the solar chromosphere are classified as (two-level) scattering lines due to their low photon destruction probabilities, such as \CaHandK, the \CaII\ infrared triplet, \MgIIHandK, the \MgIb\ triplet, the \NaId\ doublet lines, as well as the \Lybeta\ line.
For these spectral lines, the source functions follow mostly the mean radiation field $\overline{J}_{\nu}$, which suggests two-level scattering as a good approximation.
However, our results for \Lybeta\ suggest that having $S_\nu^l \approx \overline{J}_{\nu}$ is a necessary but not sufficient condition to classify a line as (two-level) scattering.
\Lybeta\ is strongly influenced by interlocking effects, formed under a ``two-level'' detailed balance, which implies the need to compare the indirect against the direct transition rates.

To classify chromospheric spectral lines based on the processes that create most of the observed line photons, namely two-level or interlocking one has to calculate the indirect transition rates to evaluate the multi-level source function.
It would be of interest to apply our multi-level source function description to classify the most important chromospheric spectral lines to get a better understanding on their formation in the solar chromosphere.

Our results on the formation of \Lybeta\ support the fact that \Lybeta\ is affected by cross-redistribution, also known as Raman scattering \citep{Hubeny1995}.
\Lybeta\ is strongly coupled with \Lyalpha\ and \Halpha\ and therefore scattering between these lines and redistribution of photons will be very effective \citep{Heinzel1985}.
As the three hydrogen lines are strongly coupled this might indicate that \Halpha\ is an interlocking line and not a two-level scattering line as suggested by \citet{Rutten2012}.
\citet{Rutten2012} did not calculate the indirect transition rates explicitly and their Fig. 12 illustrating that \Halpha\ is a two-level scattering line could be misleading if \Halpha\ is formed under ``two-level'' detailed balance with $\overline{J}_{\nu} \approx B_{\nu_0} (T^{\star})$.

We show that absorbing Markov chains can represent the level-ratio solution of the statistical equilibrium equation made up of direct and indirect transition rates. 
The indirect transition rates are the sum of the transitions per second from the lower/upper level into the individual intermediate levels times an indirect transition probability.
The indirect transition probabilities \citep[as introduced by][]{Jefferies1960} represent all non-coherent paths from the individual intermediate level to the upper/lower level of the atomic transition.
Interlocking becomes important if there is a large imbalance of indirect transitions between the upper and lower levels of a transition represented by a multi-level source function.
The multi-level source function can help interpret spectral line formation from modern multi-level non-LTE calculation as illustrated on \Lyalpha, \Lybeta, and \Lygamma.
\Lybeta\ and \Lygamma\ are strongly influenced by interlocking and formed under ``two-level'' or interlocking detailed balance that gives $\overline{J}_{\nu} \approx B_{\nu_0} (T^{\star})$. 

Formally, our method is valid only for statistical equilibrium between levels, where ionization is included as one or more levels. However, it may be possible to apply our method also in cases of non-equilibrium ionization by keeping the ionization fraction constant and solving statistical equilibrium only for the excited states of the neutral atom, as done by \cite{Krikova2023}. This approximation is valid only for model atoms with a single ionized level. 

\section{Conclusions}
\label{sec:Conclusion}

We use Markov chain theory to interpret a specific form of the multi-level source function.
In non-LTE radiative transfer, the multi-level source function is a key quantity to interpret optically thick line formation, and it depends on the level-ratio solution of the statistical equilibrium equations.
We find that absorbing Markov chains are a valid alternative approach to solving the statistical equilibrium equation in terms of level ratios.
A crucial advantage of this new method is that it can be used to quantify the effects of interlocking in multi-level atoms.

The effects of interlocking are described by the indirect transition rates, which quantify the transitions per second through intermediate levels that end up in the lower or upper level of an atomic transition. They are the sum of the transitions per second to the intermediate level times an indirect transition probability.
The absorbing Markov chain highlights that indirect transition probabilities, as introduced by \citet{Jefferies1960}, represent all paths of an atomic transition from the intermediate level leading to the upper or lower level and not entering the same level twice.
This insight into the origin of the indirect transition/probabilities rates combined with Eq. (\ref{eq:alpha_d}) gives a straightforward explanation to when interlocking becomes important.
Equations (\ref{eq:alpha_d}), (\ref{eq:msf_eta}), and (\ref{eq:msf_sf}) tell us that if the imbalance of transitions connecting the upper and lower levels through all non-coherent paths becomes greater than the scattering or thermal extinctions, the source function (and therefore the source of photons) is strongly coupled to the formation of other spectral lines at different wavelength positions.
Therefore, interlocking is always non-local in wavelength.

We present a general form of the multi-level source function (Eq. [\ref{eq:msf_sf}]) that allows one to determine which higher-order transition path dominates the interlocking source function (Eqs. [\ref{eq:interlocking_sf_split}]-[\ref{eq:interlocking_sf_omega}]).
Several previous studies relied upon a more qualitative assessment of interlocking \citep{Bruls1992, Kneer2010, Leenaarts2010, Rutten2012} and did not evaluate the indirect transition rates explicitly.

Our analysis of the formation of \Lybeta\ and \Lygamma\ highlights that a quantitative assessment of the multi-level source function is necessary to determine the physical process dominating the line source function.
By calculating the indirect transition rates we find that \Lybeta\ and \Lygamma\ can be classified as interlocking lines in the solar FALC model.
A qualitative analysis of the source function of \Lybeta\ and \Lygamma\, such as the two-level approximation or the method used by \citet{Rutten2012} can lead to a wrong classification of the source function.
As a consequence, one might draw the wrong conclusion about the formation of a spectral line.
In particular, if the line is formed under the condition of ``two-level'' or interlocking detailed balance resulting in $\overline{J}_{\nu} \approx B_{\nu_0} (T^{\star})$.
The \Halpha\ line, one of the most studied chromospheric spectral lines might be formed under such a condition.
Our analysis hints that \Halpha\ might be strongly interlocked with \Lyalpha\ and \Lybeta.
For a definitive answer about the effect of interlocking on \Halpha\ and formation, one should perform a quantitative analysis on the source function of \Halpha\ using the multi-level source function introduced in Sect. \ref{sec:markov_msf}. 
The same is true for most of the strong spectral lines formed in the chromosphere: it may be that interlocking is the dominant mechanism setting the source function, and the methodology outline here is ideally suited for such studies.
%

\begin{acknowledgements}
This work has been supported by the Research Council of Norway through its Centers of Excellence scheme, project number 262622. 
Computational resources have been provided by Sigma2 – the National Infrastructure for High-Performance Computing and Data Storage in Norway. We acknowledge funding support by the European Research Council under ERC Synergy grant agreement No. 810218 (Whole Sun).
\end{acknowledgements}

\bibliographystyle{aasjournal}
\bibliography{mybib}

\end{document}